\documentclass[a4paper,11pt]{article}
\pdfoutput=1 % if your are submitting a pdflatex (i.e., if you have
\usepackage{jcappub} 
\usepackage{graphicx}
\usepackage{braket}
\usepackage{subcaption}
\usepackage[T1]{fontenc} % if needed

\title{\boldmath The Preheating Stage on The Starobinsky Inflation after ACT}

\author[a,b]{Norma Sidik Risdianto\footnote{Corresponding author}}
\author[c]{Romy Hanang Setya Budhi}
\author[c]{Nehla Shobcha}
\author[a]{Apriadi Salim Adam}
\author[c]{Muhammad Abdan Syakura}
\affiliation[a]{Research Center for Quantum Physics, National Research and Innovation Agency (BRIN), \\
South Tangerang 15314, Indonesia}
\affiliation[b]{Dept. of Physics Education, Universitas Islam Negeri Sunan Kalijaga,\\Jl. Marsda Adisucipto 55281, Yogyakarta, Indonesia}
\affiliation[c]{Physics Department, Universitas Gadjah Mada, Yogyakarta 55281, Indonesia}

\emailAdd{norma.risdianto@uin-suka.ac.id}
\emailAdd{romyhanang@ugm.ac.id}
\emailAdd{nehlashobcha@mail.ugm.ac.id}
\emailAdd{apriadi.salim.adam@brin.go.id}
\emailAdd{muhammadabdansyakura1997@mail.ugm.ac.id}

\abstract{Recent ACT results favor larger e-fold numbers, which tighten the reheating requirements in Starobinsky inflation. We show that efficient reheating generically requires an additional spectator-assisted preheating stage. The preheating dynamics and reheating temperature must be significantly modified to accommodate the higher reheating temperature. We present viable parameter sets compatible with PBH formation and the preferred reheating mechanism in this model. Furthermore, the momentum spectrum evolves from infrared-dominated modes at the early stage of preheating to higher-momentum modes at later times, supporting our scenario.
}

\begin{document}
\maketitle
\flushbottom
\newpage

\section{Introduction}

The Starobinsky model \cite{starobinsky}, as the simplest example of modified gravity \cite{de2010f, sotiriou2010f, nojiri2011unified}, is considered one of the most successful inflationary models. Its simplicity lies in the fact that it relies solely on gravity as the driving mechanism for inflation. Despite this simplicity, it remains consistent with the Cosmic Microwave Background (CMB) observations from PLANCK data \cite{cmb}. In recent years, this model and its various extensions have continued to attract considerable attention and discussion (see e.g. refs. \cite{kehagias, Ketov:2025nkr}). However, the Starobinsky model is now challenged by newly established observational data from the Atacama Cosmology Telescope (ACT) \cite{louis2025atacama}. This new data shows that the slow-roll parameter is  $n_s = 0.9709 \pm 0.0038$. When combined with measurements from CMB lensing, Baryon Acoustic Oscillation (BAO), and the Dark Energy Spectroscopic Instrument (DESI), the extended result yields $n_s = 0.9743 \pm 0.0034$. This result indicates that the Starobinsky model is excluded at the $2\sigma$ level for $N\approx 60$. The model remains viable if a large number of e-folds is allowed \cite{Drees:2025ngb,liu2025reconciling} (see also refs. \cite{mohammadi2025starobinsky, zharov2025reheating}). Alternatively, this issue can be addressed by introducing higher-curvature perturbations to satisfy the new constraint \cite{addazi2025curvature}.

 The reheating temperature of the Starobinsky model is found to be relatively low $\sim 10^9$ GeV \cite{de2010f,dorsch2024gravitational}\footnote{See also $f(R)$ for the constant roll mode, which can alter the reheating temperature \cite{oikonomou2017reheating}. It is also noted that the reheating temperature on $f(R)$ is also discussed in ref. \cite{odintsov2025power,Odintsov:2026doe} and \cite{oikonomou2025strong}} compared to other models, such as Higgs inflation \cite{bezrukov2009initial}. This lower reheating temperature is often preferred in inflationary cosmology, as it helps avoid the overproduction of unwanted relics \cite{kallosh2000gravitino}. For example, Refs. \cite{ellis1982inflation,kawasaki1995gravitino} argue\footnote{See also ref. \cite{mcdonald2000reheating}.} that the reheating temperature should be below $10^9$ GeV to prevent excessive production of gravitino.
Such a low reheating temperature significantly influences the thermal history of the universe, particularly with respect to baryogenesis \cite{davidson2000baryogenesis} and the production of dark matter. However, as for leptogenesis, it requires at least several MeV of reheating temperatures to operate successfully \cite{hannestad2004lowest,choi2017new}.
In summary, estimation of the reheating temperature range is wide, from a few MeV up to nearly the Planck scale. This broad range suggests that the reheating temperature is still poorly constrained. In addition to the ACT observation results, the Starobinsky inflation model is expected to yield a much higher reheating temperature.

The preheating stage of the Starobinsky model has received comparatively less attention in the literature than its inflationary stage. The explanation should be started as the fact that inflationary dynamics are better constrained, whereas the preheating phase tends to be strongly model-dependent \cite{van2017reheating,fu2019nonlinear}.\footnote{One can see the historical preheating stage in Refs. \cite{Dolgov:1989us,Traschen:1990sw} (see also refs. \cite{Shtanov:1994ce,Kofman:1994rk})} In the Starobinsky model, preheating is primarily driven by the interaction of the produced particles with gravity. While this scenario is simple and potentially promising, our limited understanding of gravity makes it more challenging to be favored as a realistic theory. In this paper, we investigate the preheating mechanism in the Starobinsky model, focusing on the efficient energy drain through preheating, the constraints imposed by primordial black hole (PBH) production, and the resulting reheating mechanism.

This paper is organized as follows: In Section \ref{model}, we discuss the inflationary description of the Starobinsky model. In Section \ref{secreheating}, we estimate the resulting reheating temperature by the ACT constraint.  In Section \ref{preheating}, we examine the preheating stage of the Starobinsky model, which is non-minimally coupled to scalars. In Section \ref{fermionicpreheating}, we consider the possibility of fermionic preheating in our model. We also discuss primordial black hole (PBH) formation in Section \ref{pbh}. The realistic reheating scenario is discussed in Section \ref{reheatingscenario}. Finally, we present our conclusions in Section \ref{conclusion}.

\section{The inflationary model}\label{model}
We start this section by writing the action of the model in the Jordan frame as\footnote{In this paper, we used $(-,+,+,+)$ convention.}
\begin{equation}
    S_J=\int d^4x\sqrt{-g_J}\left[\frac{1}{2}M_p^2f(R_J) \right], \hspace{1cm} f(R_J)= R_J+\frac{R^2_J}{6M^2},
\end{equation}
where $g_J$ is the determinant of the metric tensor $g^{\mu\nu}_J$, $M_p=1/\sqrt{8\pi G}$ is the reduced Planck mass, $R_J$ is the Ricci scalar, and $M$ is the effective inflaton's mass at the end of inflation. The subscript $J$ represents the Jordan frame. By using the conformal transformation as
\begin{equation}
    g^{\mu\nu}_E\equiv\Omega^{-2}  g^{\mu\nu}_J, \hspace{1cm}\Omega^{2}\equiv\frac{df(R_J)}{dR_J}=1+\frac{R_J}{3M^2}\equiv\exp\left({\sqrt{\frac{2}{3}}\frac{\phi}{M_p}}\right),
\end{equation}
we obtain the action in the Einstein frame as follows
\begin{equation}\label{se}
    S_E=\int d^4x \sqrt{-g_E}\left[\frac{1}{2}M_p^2 R_E -\frac{1}{2}\partial^\mu \phi\partial_\mu \phi-V(\phi) \right],
\end{equation}
where ${R}_E=R_J\Omega^{-2}-6\Omega^{-3}\square\Omega$ and

\begin{equation}\label{potentialphi}
    V(\phi)=\frac{3}{4}M_p^2M^2\left(1-e^{-\sqrt{\frac{2}{3}}\frac{\phi}{M_p}}\right)^2.
\end{equation}For the rest of this paper, we will omit the subscript $E$ for simplicity. Also, Throughout this work, we assume a spatially flat FRW metric in the Einstein frame after the conformal transformation.

Firstly, we will write the scalar spectral index ($n_s$), tensor-to-scalar ratio ($r$), slow-roll parameters ($\epsilon$ and $\eta$) and the scalar power spectrum ($A_s$) respectively as

\begin{equation}\label{ns}
    n_s = 1 - 6\epsilon + 2\eta, \quad r = 16\epsilon, \quad \epsilon=\frac{1}{2}M_p^2 \left(\frac{V'}{V} \right)^2, \quad \eta= M_p^2\frac{V''}{V}, \quad A_s = \frac{V(\phi)}{24\pi^2 \epsilon} .
\end{equation}
The prime denotes the derivative with respect to scalaron $\phi$. By using the ACT from the extended dataset P-ACT-LB yields \cite{louis2025atacama}  $n_s = 0.9743 \pm 0.0038$, $A_s \approx 2.13 \times 10^{-9}$ at the pivot scale $\kappa_*=0.05$ Mpc$^{-1}$. The $n_s$ can be used to determine the new constraint on the $\phi_\text{ini}$, which corresponds to the field value of the scalaron at $\kappa_*$, to be 
\begin{equation}
    \phi_\text{ini}\approx 5.72\hspace{1mm} M_p,
\end{equation}
where this value is set by the center value of $n_s$, and the details may be explained shortly. The inflaton's field value at the end of inflation is obtained by setting $\epsilon=1$, which corresponds to the violation of the slow-roll conditions as
\begin{equation}
    \phi_\text{end}\simeq 0.94 \hspace{1mm}M_p.
\end{equation}
By using the last two results, we can obtain the number of e-folds $N$ as
\begin{equation}\label{efold0}
    N=\frac{1}{M_p^2}\int^{\phi_\text{ini}}_{\phi_\text{end}}\frac{V}{V'}d\phi\simeq 75.51,
\end{equation}
much higher than the previously predicted by PLANCK results. It means that the increase in $n_s$ slightly by ACT from the 
extended dataset P-ACT-LB  results could increase the e-fold greatly.\footnote{The choice of an e-fold number greater than 60 is rather implausible, although it remains acceptable \cite{liddle2003long}.} Note that we only use the central value $n_s=0.9743$, which gives an e-fold number of $N=75.5$, as obtained from Eqs. \eqref{ns}$-$\eqref{efold0}. In addition, using the full range of $n_s$ yields an e-fold range of $66\text{--}87$, which significantly alters the reheating temperature. One should note that the discussed value of $n_s$ lies within the $1\sigma$ range. Consequently, such a large e-fold number may lead to a dangerously high reheating temperature beyond the Planck scale (see Section \ref{secreheating}). In the next section, we show that Starobinsky inflation is no longer favored under the $1\sigma$ constraint, although it still remains viable within the $2\sigma$ range.\footnote{Strictly speaking, $N=60$ lies at the edge of the $2\sigma$ range. Therefore, Starobinsky inflation may already be in a marginally disfavored region.} The effective scalaron's mass has been slightly altered by ACT results as $M=9.07\times 10^{-6}M_p$.\footnote{The scalaron mass M is determined from the normalization of the scalar power spectrum using the ACT best-fit values. The model is then tested against the ACT confidence regions, where the conventional Starobinsky model remains viable only within the $2\sigma$ allowed region.}

\section{The reheating constraint by ACT results}\label{secreheating}

In this paper, we will estimate the reheating temperature with the new ACT results. This constrained reheating temperature will be important to estimate the proper reheating process as depicted in section \ref{reheatingscenario}. However, the calculation depicted in Eq. \eqref{efold} will be conventional. 
The reheating temperature will be evaluated by using 

\begin{equation}\label{efold}
\frac{\kappa_*}{a_0 H_0}= \frac{a_\kappa}{a_\text{end}}\frac{a_\text{end}}{a_\text{reh}}\frac{a_\text{reh}}{a_\text{eq}}\frac{a_\text{eq}}{a_0}\frac{H_\text{eq}}{H_0}\frac{H_\kappa}{H_\text{eq}},
\end{equation}
where $a_0=1$ and $H_0=66.78 \hspace{1mm}\text{km Mpc}^{-1}\text{s}^{-1}$ \cite{louis2025atacama} represent the scale factor and Hubble parameter today. $a_\kappa$ and $H_\kappa$ represent the scale factor and Hubble parameter at the pivot scale $\kappa_*=0.05 \hspace{1mm}\text{Mpc}^{-1}$. The subscripts 'eq', 'end', and 'reh' correspond to  'matter-radiation equality', 'end of inflation', and 'reheating', respectively. Note that we solve Eq. \eqref{efold} by using $T_\text{eq}=0.8$ eV. On the other hand, we used $\frac{a_\text{eq}H_\text{eq}}{a_0H_0}\simeq\sqrt{1+z_\text{eq}}\simeq 3400$ by assuming the matter domination near equality. Straightforwardly, an e-fold number of $N=55\sim56$ can produce $T_\text{reh}=10^{13}\sim10^{15}$ GeV. This range is still favored by ACT-LB at the edge of $2\sigma$ for $w\simeq 0$. However, these parameter choices are excluded at the $2\sigma$ level by the P-ACT-LB results. The reason for choosing these values will become clear shortly.

Recent P-ACT-LB results have suggested that the effective post-inflationary equation of state may favor values $w\gtrsim 0.5$ \cite{zharov2025reheating}. Similar conclusions have also been obtained in refs. \cite{Drees:2025ngb,mohammadi2025starobinsky}. 
In this work, we focus on this surviving parameter space and adopt $w=0 \sim 0.1$ to describe the early post-inflationary stage. As stated above, this choice lies at the edge of the ACT-LB constraint at the $2\sigma$ level. It is motivated by the scalaron oscillating around a quadratic minimum, $V(\phi)\propto\phi^2$, which naturally yields an averaged equation of state $w\simeq0$. Moreover, the preheating mechanism studied in this paper relies on the efficient production of heavy and initially non-relativistic particles, a process that is naturally realized during a matter-dominated phase.
Straightforwardly, our approach differs from that of ref. \cite{Drees:2025ngb}, which focuses primarily on the observational constraints on the reheating history rather than on constructing a specific preheating mechanism.

One should note that even a small change in $n_s$ can lead to a substantial variation in the number of e-folds and, consequently, a significant change in the reheating temperature. In this work, we argue that the reheating temperature is directly connected to the preheating stage. As a result, the following sections explain how both the preheating stage and the reheating mechanism are adjusted to accommodate the new, higher reheating temperature favored by ACT results.\footnote{One should note that the deformation of $f(R)$ theory could also alter the reheating temperature (See e.g. ref. \cite{odintsov2025power,oikonomou2025strong,Odintsov:2026doe}). See also ref. \cite{Oikonomou:2026mvp} for the reheating temperature in the single-field case, similar to our paper.}

\section{The preheating}\label{preheating}
When $\Tilde{\phi}\lesssim M_P$,  the inflaton oscillates, marking the end of inflation. The particle produced by this mode is non-perturbative. If we assume $\chi$ is dominantly produced during this mode, which is presented by the action in the Jordan frame as:
\begin{equation}
    S_\chi=\int d^4x \sqrt{-g_J}\left[-\frac{1}{2}g^{\mu\nu}_J\partial_\mu \chi \partial_\nu \chi -\frac{1}{2} m_\chi^2 \chi^2 -\frac{1}{2}\xi R_J \chi^2\right],
\end{equation}
we can evaluate its particle production. 
Before we proceed, we need to clarify that after the end of inflation, the potential \eqref{potentialphi} becomes quadratic and the analytic solution of $\phi$ can be approximated as  \cite{kofman1997towards}

\begin{figure}[t]
    \centering
    \includegraphics[width=0.6\linewidth]{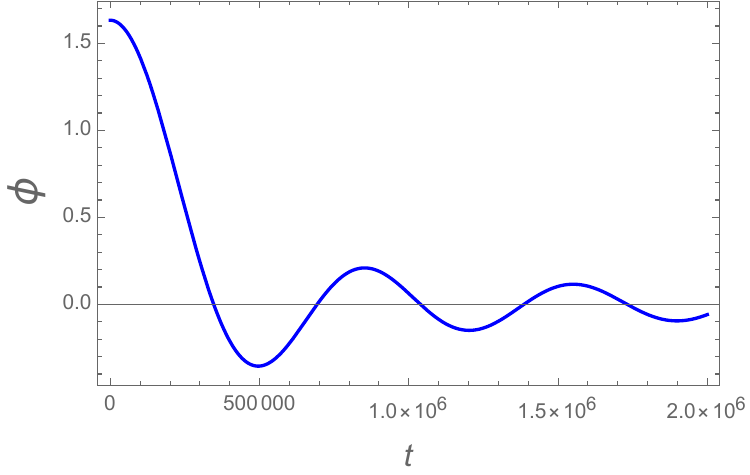}
    \caption{The plot of $\phi$ with time $t$. The unit of $\phi$ is $M_p$ while the unit of $t$ is in $M_p^{-1}$.}
    \label{eomphi}
\end{figure}

\begin{equation}\label{phi}
    \phi(t)=\Tilde{\phi}(t) \sin (M t), \hspace{1cm} \Tilde{\phi}(t)=2\sqrt{\frac{2}{3}} \frac{M_p}{Mt}.
\end{equation}
Equation~\eqref{phi} is employed as an analytical approximation for the oscillating scalaron background during preheating. We do not assume that this solution remains valid throughout the entire preheating stage. Near the end of preheating, backreaction effects should be taken into account. However, doing so would require solving coupled nonlinear equations of motion and possibly performing lattice simulations. These treatments are beyond the scope of this paper and are left for future work. We emphasize that our main conclusions do not rely on the precise late-time evolution of the scalaron field. Thus, this approximation remains valid for obtaining the favored reheating temperature, which is the main goal of this paper.

\begin{figure}[t]
    \centering
    \includegraphics[width=1\linewidth]{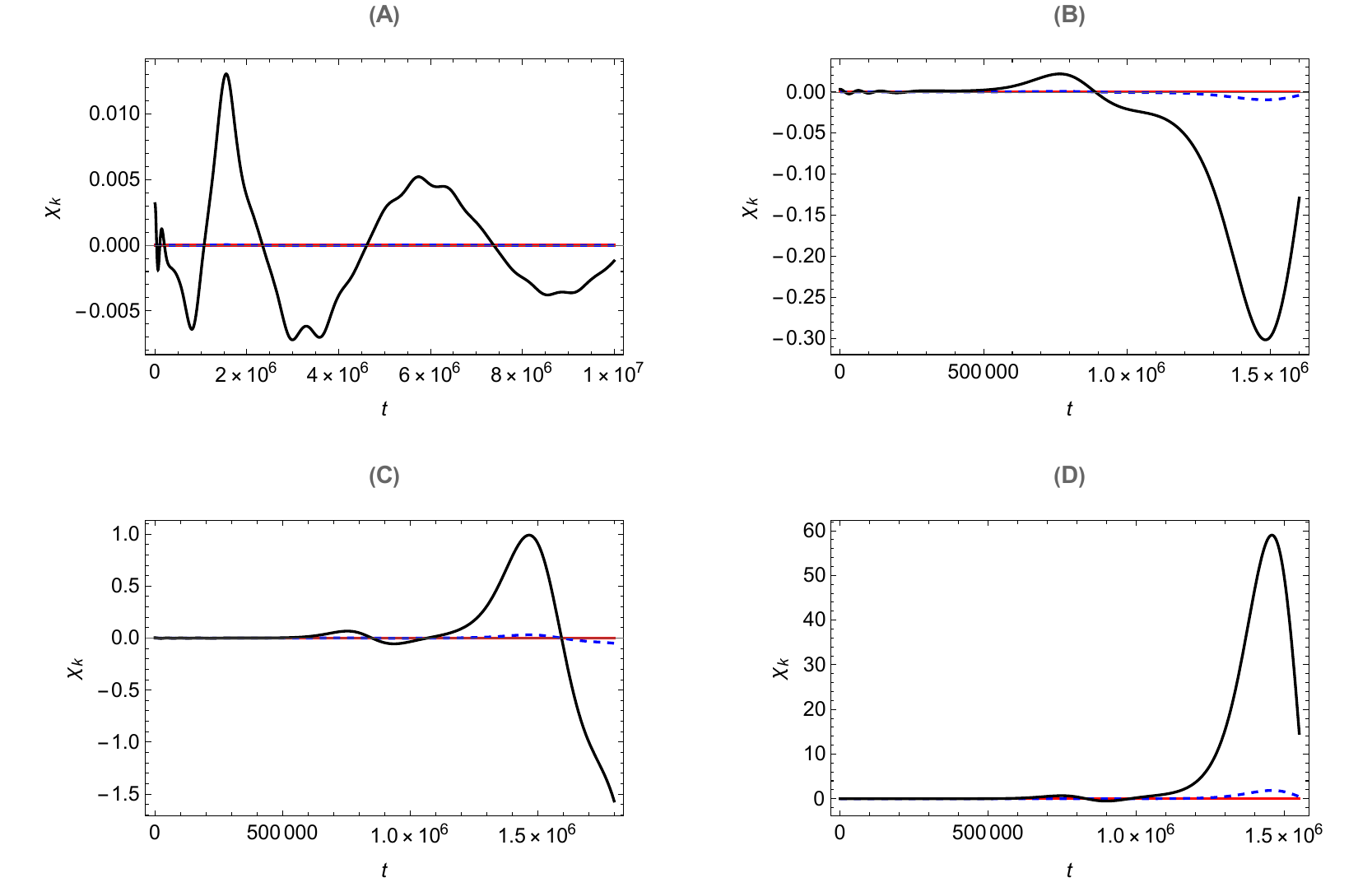}
    \caption{The growth of $\chi_k$ by the resonance production. We fixed $M=9.07 \times 10^{-6}M_p$, $M_p=1$, $k_\chi=10^{-8}M_p$, $m_\chi=10^{-8}M_p$, and $a=1$. The solid-red represents to the $\chi_k(0)=0$, the dashed-blue represents to the $\chi_k(0)=10^{-5}M_p$, and the solid-black represents to the $\chi_k(0)=10^{-2.5}M_p$. The varied $\xi$ is depicted in the corresponding panels: panel (A) uses $\xi=1$, panel (B) uses $\xi=4$, panel (C) uses $\xi=7$, and panel (D) uses $\xi=10$. $t$ is in a unit of $  M_p^{-1}$. In all numerical calculations in this figure, we used $\dot{\chi}_k(0)=0$. }
    \label{kksmall}
\end{figure}

Furthermore, by transforming $S_\chi$ in the Einstein frame as $S_{\chi, E}$, the equation of motion (e.o.m.) of $\chi$ can be written by varying $S_E+S_{\chi,E}$ by $\delta\chi$  as

\begin{equation}\label{eomchi}
 \Ddot{\chi}_k+3H\Dot{\chi}_k-\sqrt{\frac{2}{3}}\frac{\Dot{\phi}\Dot{\chi}_k}{M_p}+\left[\frac{k^2_\chi}{a^2}+e^{-\sqrt{\frac{2}{3}}\frac{\phi}{M_p}}m_\chi^2 + 3\xi M^2  \left(1-e^{-\sqrt{\frac{2}{3}}\frac{\phi}{M_p}}\right) \right] \chi_k=0.
\end{equation}
where $H$ is the Hubble parameter, and we used the Heisenberg representation as 
\begin{equation}\label{heisenberg}
    \chi(\textbf{x},t)=\int \frac{d^3k}{(2\pi)^{\frac{3}{2}}}\left(\hat{a}_k \chi_{k}(t)e^{-i\textbf{k}_\chi\cdot \textbf{x}} +\hat{a}^\dagger_k \chi^*_{k}(t)e^{i\textbf{k}_\chi\cdot \textbf{x}}\right),
\end{equation}
on calculating \eqref{eomchi}. Note that the term $\sqrt{\frac{2}{3}}\frac{\Dot{\phi}\Dot{\chi}_k}{M_p}$ produces the anomalous friction term and the non-canonically normalized $\chi_k$. This problem can be easily solved by shifting $\chi_k\rightarrow e^{\frac{1}{\sqrt{6}}\frac{\phi}{M_p}}\chi_k$. The transformation is introduced to canonically normalize the kinetic term and eliminate the anomalous friction term and the physical particle production is unchanged. Finally, we obtain

\begin{equation}\label{ddotchi}
    \Ddot{\chi}_k+3H\Dot{\chi}_k+\left[\frac{k^2_\chi}{a^2}   +e^{-\sqrt{\frac{2}{3}}\frac{\phi}{M_p}}m_\chi^2 + 3\xi M^2  \left(1-e^{-\sqrt{\frac{2}{3}}\frac{\phi}{M_p}} \right) -\frac{\Dot{\phi}^2}{6M_p^2}+\frac{\Ddot{\phi}+3H\Dot{\phi}}{\sqrt{6}M_p}\right]\chi_k =0
\end{equation}
In Fig. \ref{kksmall}, we show that the resonance production of $\chi_k$ by solving Eq. \eqref{ddotchi} is highly sensitive to both the non-minimal coupling $\xi$ and the initial condition $\chi_k(0)$. If both $\xi$ and $\chi_k(0)$ are not sufficient, Starobinsky inflation faces two possible results: (1) inconsistency with the observed particle abundance today (especially if $\chi$ is related to dark matter) or (2) prolonged preheating to generate the required number of particles. In addition, as discussed later in Sections \ref{pbh} and \ref{reheatingscenario}, the PBH and reheating temperature are less sensitive to the efficiency of the early preheating stage. Therefore, the effects of the  the non-minimal coupling $\xi$ and  the initial conditions are not significant. 

The initial conditions can be provided via four mechanisms: perturbative decay at an early stage, scattering of the scalaron, quantum fluctuation $\delta \chi$, and $\chi$ as a spectator field during inflation. Among the four mechanisms, we will qualitatively analyze which mechanism is the most favored.

At the early stage of preheating, the inflaton perturbative decay and the scattering process are possible. The single perturbative decay, in principle, is more favored than scattering, as it needs a large density of incoherent inflaton particles to scatter. In contrast, at the start of preheating, the inflaton is mostly a coherent oscillating field. The perturbative decay itself is slow to be produced at the early stage of preheating. Additionally, if the excessive decay occurs during inflation, it may spoil the slow-roll condition or leave observational relics. On the other hand, the favored $\chi\sim 10^{-2.5}M_p$ could be obtained from the fluctuation of $\delta\chi$. But the fluctuation can only provide as large as $H/2\pi\sim 10^{-6}M_p$, which is about three orders smaller than the desired value. Lastly, the only possible way came from the fact that $\chi$ should be the spectator field during inflation. Straightforwardly, in this paper we argue that the presence of a spectator field is necessary for successful preheating in Starobinsky inflation. At the early stage of preheating, satisfying both the non-relativistic condition and efficient preheating requires $k_\chi$ to be extremely small. Consequently, this leads to the condition $k_\chi^2 \ll \xi M^2$.

Before proceeding, we clarify why the analytical time evolution of $\phi$ is used, even though it could in principle be numerically coupled with $\chi$ to obtain exact results. \textit{First}, in the pure Starobinsky model, the scalaron $\phi$ is only weakly coupled to other fields during the initial stage of preheating. In the Einstein frame, the Lagrangian of the $\chi$ field is heavily suppressed by the factor $\exp\left(-\sqrt{2/3} \phi/M_p\right)$, which indicates that $\phi$ and $\chi$ are weakly coupled. This justifies treating $\chi$ as a spectator field during inflation, so that the analytical form of Eq. \eqref{phi} remains valid until $\phi$ becomes small. Also, we restrict our attention to the early phase, where $\phi$ is still not too small after the first crossing. \textit{Second}, solving the coupled equations of motion poses a significant challenge, as it involves complicated calculations, including integro-differential equations in $\chi^2$. \textit{Third}, since our paper primarily focuses on the reheating mechanism in light of the new ACT results, we approximate the preheating by considering how efficiently it can drain the inflaton field energy density. A more detailed analysis of the preheating stage may be presented in future work.

\begin{figure}
    \centering
    \includegraphics[width=0.5\linewidth]{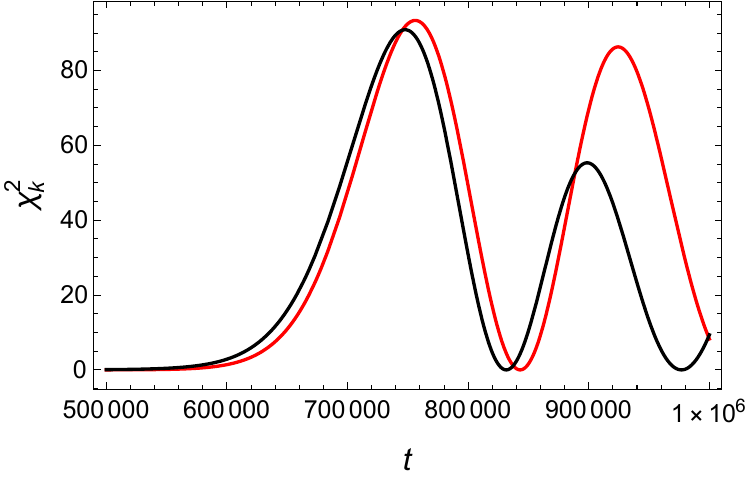}
    \caption{The comparison of analytical result (solid-red) and numerical result (solid-black) growths in $\chi_k^2$. The numerical result has been magnified by $200$ times to match the analytical ones. In this plot, we used $\xi=10$, $k_\chi=10^{-8}M_p$, $m_\chi=10^{-8}M_p$, $\chi_k(0)=10^{2.5}M_p$ and $\Dot{\chi}_k=0$. The unit of $t$-axis is in $M_p^{-1}$.}
    \label{analytic-numeric}
\end{figure}

In the following, we will describe the analytical approximation to dig further into the particle production of $\chi_k$. At the end of inflation, Eq. \eqref{ddotchi} can be approximated by using $a=1$ and evaluated when $\Tilde{\phi}<M_p$ as

\begin{equation}\label{chioscillation}
    \Ddot{\chi}_k+\Bigg[ (k^2_\chi+m_\chi^2) + \frac{ M^2  \sqrt{6}}{M_p}\left(\xi-\frac{1}{6}-\frac{m_\chi^2}{M^2}\right)\tilde{\phi}\sin(Mt)\Bigg] \chi_k= 0,  
\end{equation}
where we used $\Ddot{\phi}+3H\Dot{\phi}=-\frac{dV}{d\phi}$ and assumed $m_\chi\ll M$. Note that the conformal coupling corresponds to $\xi=1/6$ \cite{de2010f}. The approximation adopted in Eq. \eqref{chioscillation} is likely oversimplified. As illustrated in Fig. \ref{analytic-numeric}, the analytical solution shows a significant deviation from the full equation of motion given in Eq. \eqref{ddotchi}. In that figure, we present a comparison of $\chi_k^2$ rather than $\chi_k$ itself, since it directly reflects the produced energy density $\delta \rho_\text{cross} \propto |\chi_k|^2$. Figure \ref{analytic-numeric} shows that the numerically obtained energy density of the $\chi$ field is approximately two orders lower than the analytical estimation. We will use these results to approximate the particle production during the crossing from the analytical calculations.

To estimate the analytical assumption, we need to focus on the particle production during the first near-zero crossing ($\sin(Mt)\simeq Mt$). Firstly, we can approximate Eq. \eqref{chioscillation} as
\begin{equation}\label{chioscillation2}
    \frac{d^2{\chi}_k}{dq^2}+\left(p^2 +q\right)\chi_k=0,
\end{equation}
where
\begin{equation}\label{chioscillation3}
\begin{split}
    & p^2=\left(k_\chi^2+m_\chi^2\right) \left[\frac{ M^3  \sqrt{6}}{M_p}\left(\xi-\frac{1}{6}-\frac{m_\chi^2}{3M^2}\right)\tilde{\phi}\right]^{-2/3}, \\
    &q=\left[\frac{ M^3  \sqrt{6}}{M_p}\left(\xi-\frac{1}{6}-\frac{m_\chi^2}{3M^2}\right)\tilde{\phi} \right]^{1/3}t.
\end{split}
\end{equation}
This simplification will help us to understand the numerical results depicted in Fig. \ref{kksmall}, which is obtained by solving Eq. \eqref{ddotchi}. 

For $p^2\ll q$, we can estimate the particle production of $\chi$ by the first crossing of $\phi$ as \cite{bezrukov2009initial,risdianto2025second}
\begin{equation}\label{deltarho}
    \delta \rho = \int^\infty_0 \frac{d^3k_\chi}{(2\pi)^3}e^{-\pi p^2} \Bar{m}
\end{equation}
where
\begin{equation}\label{barmass}
\Bar{m}= \sqrt{  \frac{ M^2  \sqrt{6}}{M_p}\left(\xi-\frac{1}{6}-\frac{m_\chi^2}{3M^2}\right)  \Tilde{\phi} }
\end{equation}

Analytically, by calculating Eq. \eqref{deltarho}, we obtain that in the first zero crossing (after the end of inflation), it produces
\begin{equation}
    \delta \rho_\text{cross} \simeq  0.27 \times \xi^{3/2} M^4\sim 10^{-21} \xi^{3/2} M_p^4,
\end{equation}
if we set $\xi>1$.
In this calculation, for the crossing to achieve the same energy as the inflaton ($\tfrac{1}{2}M^2 \Tilde{\phi}^2 \sim 10^{-11}M_p^4$), it requires $\xi \gg \mathcal{O}(1)$, which is strongly disfavored due to naturalness issues. In addition, In the caption of Fig. \ref{analytic-numeric} shows that the actual value of $\delta \rho_\text{cross}$ obtained numerically is about two orders smaller than of the analytical results. These results are largely due to our assumption that $a=1$. Thus, our results shown that the particle production in the Starobinsky model during matter-dominated regime is inefficient.

\begin{figure}[t]
    \centering
    \includegraphics[width=0.6\linewidth]{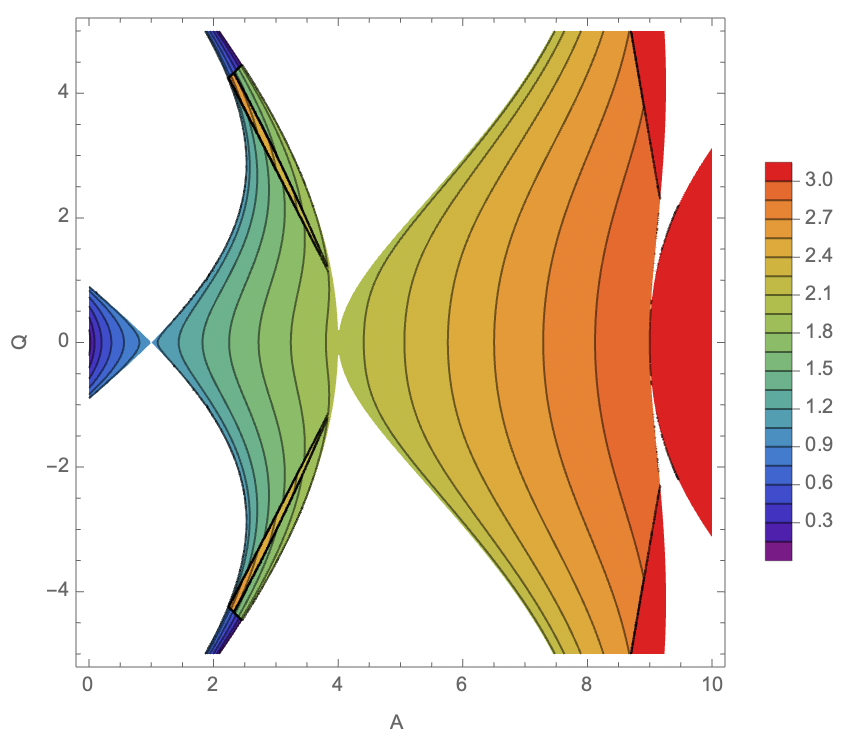}
    \caption{Mathieu instability region related to the semi-analytical result depicted in Eq. \eqref{mathieu}. The right legend corresponds to the value of $\mu_p$. The white region corresponds to the \textit{stable} condition.}
    \label{mathieuplot}
\end{figure}

Lastly, we will describe the e.o.m. of $\chi_k$ in the redefined analytical result. Straightforwardly,  Eq. \eqref{chioscillation} can be shown by
\begin{equation}\label{mathieuparameters}
\begin{split}
       &Mt\equiv 2z-\pi/2, \hspace{1cm} A\equiv 4\left(\frac{k_\chi^2+m_\chi^2}{M^2} \right),\\
       &Q\equiv\frac{ 2\sqrt{6}}{M_p}\left(\xi-\frac{1}{6}-\frac{m_\chi^2}{3M^2}\right)\tilde{\phi}.
\end{split}
\end{equation} 
Finally,  we obtain the famous Mathieu equation as

\begin{equation}\label{mathieu}
    \frac{d^2\chi_k}{dz^2}+(A-2Q\cos(2z))\chi_k=0.
\end{equation}
The numerical calculation of the Mathieu instability is shown in Fig. \ref{mathieuplot}.  The growth of the $\chi_k$ is depicted by the characteristic exponent $\mu_p$, which can be written analytically as
\begin{equation}\label{characteristicexponent}
    \mu_p=\frac{1}{2\pi}\ln\left(1+2e^{-\pi p^2}-2\sin\theta_\mu e^{-\frac{\pi p^2}{2} }\sqrt{1+2e^{-\pi p^2}}\right),
\end{equation}
where $\theta_\mu$ is a phase.
It is shown in Fig. \ref{mathieuplot} that certain parameter fall within the instability region, where the particle is largely produced. However, this semi-analytical result of the instability band could not show the effect of the initial conditions. Moreover, the parameters used in the numerical calculation can be mapped onto Fig. \ref{mathieuplot}, with some degree of deviation. Additionally, if we substitute the chosen parameters into Eq. \eqref{mathieuparameters}, most of them fall within the stability region.

\section{On the fermionic preheating}\label{fermionicpreheating}

In this part, we will discuss the possibility that, instead of the scalar $\chi$, the scalaron is coupled with fermions. So far, we have only discussed the bosonic preheating, even though fermionic preheating is equally possible. This type of preheating has already been discussed in, e.g., Refs. \cite{greene2000fermionic,greene1999fermion,giudice1999fermion,peloso2000fermion}. In this section, we will qualitatively discuss the possibility of this type of preheating.

If the scalaron is only coupled with certain fermions,  at a large coupling \cite{greene1999fermion}, the preheating is similar to the bosonic case. However, in our paper, it is heavily suppressed by the Pauli exclusion principle, in contrast to the Bose-Einstein condensate (BEC) that can occur during bosonic preheating. Fermion production becomes less suppressed once the universe expands. However, even after the end of preheating, the potential of the scalaron remains quadratic ($\propto \phi^2$), which still favors a matter-dominated regime. Fermions are more likely to be produced at the end of the preheating phase due to the looseness of the Pauli blocking.
Finally, we safely said that preheating via this mode is strongly disfavored.

\section{The PBH formation from preheating}\label{pbh}

In this section, we will discuss the PBH formation due to preheating in the Starobinsky model. We will show later that the PBH formation is consistent with it being produced during the early stage of preheating.

Straightforwardly, the constraint on the PBH can be written as
\begin{equation}\label{beta}
    \beta_\text{PBH}=\frac{\rho_\text{pbh}}{\rho_{\text{tot}}}=\int^\infty_{\delta_c} P(\delta)d\delta,
\end{equation}
where $P(\delta)$ represents the probability distribution function, and it can be written as
\begin{equation}
    P(\delta)=\frac{1}{\sqrt{2\pi} \sigma_\delta} \exp\left( -\frac{\delta^2}{2\sigma_\delta^2}\right).
\end{equation}
By using $\beta_\text{PBH}=10^{-20}$ and $\delta=\delta_c=0.7$ \cite{green2001primordial,niemeyer1999dynamics} on Eq. \eqref{beta}, we obtain the mass variance $\sigma_\delta$ as
\begin{equation}\label{sigmadelta008}
    \sigma_\delta\simeq 0.08.
\end{equation}
However, we need to skip this part before putting the constraint into our parameters.

For the production of PBH during preheating, the strong parametric resonance should be applied. Our assumptions implied that PBH is most likely produced when \cite{liddle2000super,green2001primordial}
\begin{equation}
    |Q|\equiv\left|\frac{ 2\sqrt{6}}{M_p}\left(\xi-\frac{1}{6}-\frac{m_\chi^2}{3M^2}\right)\tilde{\phi}\right| \gg A.
\end{equation}
The power spectrum of the variable $x$ can be expressed as \cite{liddle2000cosmological}
\begin{equation}
    \mathcal{P}_{(x)}=\frac{k_x^3}{2\pi^2}\braket{|x_k|^3},
\end{equation}
where $k_x = |\textbf{k}_x|$ represents the comoving wavenumber associated with $x$. Note that $x_k$ denotes the inverse Fourier transform of $x$. The influence of preheating on $\chi$ production can be determined as
\cite{kofman1997towards,liddle2000super,green2001primordial}
\begin{equation}
    \mathcal{P}_{\delta \chi}= \mathcal{P}_{\delta \chi}|_\text{end}\exp\left(2\mu_p M \Delta t\right),
\end{equation}
where $\Delta t$ corresponds to the time required to obtain the appropriate number of PBH. The characteristic exponent $\mu_p$ here is evaluated at $\theta_\mu=0$, in which $\mu_p=(2\pi)^{-1}\ln(1+2\exp(-\pi p^2))$, where $p$ is taken from Eq. \eqref{chioscillation3} as
\begin{equation}
    p^2=\frac{k_\chi^2+m_\chi^2}{\left[\frac{ M^3  \sqrt{6}}{M_p}\left(\xi-\frac{1}{6}-\frac{m_\chi^2}{3M^2}\right)\tilde{\phi}\right]^{2/3}}=\frac{1}{18 \sqrt{Q}}\left( \frac{k_\chi^2+m_\chi^2}{k_\text{end}^2}\right),
\end{equation}
where we introduce 
\begin{equation}
    k_\text{end}^2=\frac{M^2}{18\sqrt{2}}\left|\left(\xi-\frac{1}{6}-\frac{m_\chi^2}{3M^2}\right)\frac{\tilde{\phi}}{M_p}\right|^{1/6}
\end{equation}
as the comoving wavenumber evaluated at the Hubble exit at the end of inflation. For $Q\gg 1$, in which it is required to get the largest production, we get $\mu_p=\mu_0\simeq \frac{1}{2\pi}\ln 3$. In addition, we will calculate the non-adiabatic power spectrum of $\chi$ as \cite{liddle2000super,green2001primordial}
\begin{equation}
    \mathcal{P}_{\zeta_\text{n-ad}}(k_\chi)\simeq \mathcal{A}\left( \frac{k_\chi}{k_\text{end}}\right)^3 I(p, M \Delta t),
\end{equation}
where\footnote{The third bracket is coming from the amplitude of the interaction coupling between $\phi$ and $\chi$ via $\mathcal{L}_\text{int}=\sqrt{\frac{2}{3}}\frac{m^2_\chi}{M_P}\phi\chi^2$ divided by mass scale $M$.}
\begin{equation}
    \mathcal{A}\equiv\frac{3 \times 2^{3/2}}{\pi^6 \mu_p^2}\left(\frac{\Tilde{\phi}}{M_p} \right)^2\left(\frac{H_\text{end}}{M}\right)^4 \left(\sqrt{\frac{2}{3}}\frac{m^2_\chi}{M M_P}\right)^2 Q^{-1/4}
\end{equation}
and
\begin{equation}
        I(p, M \Delta t)\equiv \frac{3}{2}\int^\infty_0 p'^2dp'\int^\pi_0 \sin\vartheta d\vartheta e^{2(\mu_{p'}+\mu_{p-p'})M\Delta t}.
\end{equation}
Note that $\vartheta$ is the angle between $p$ and $p'$. To solve the last equation, we need to impose \cite{green2001primordial,liddle2000super}
\begin{equation}
    \mu_{p-p'}\simeq \mu_p'+\frac{2p'\cos \vartheta}{2+e^{\pi p'^2}}+\mathcal{O}(p^2), \hspace{1cm}\mu_{p'} \approx \mu_0+\frac{p'^2}{3}.
\end{equation}
Straightforwardly, we obtain
\begin{equation}
\begin{split}
       I(p, M\Delta t)&\simeq\frac{3}{2}\int^{\infty}_0 p'^2dp'\int^\pi_0 \sin\vartheta d\vartheta  e^{4\mu_{p'}M\Delta t}\\
       &\simeq 0.863(M\Delta t)^{-3/2}\cdot e^{4\mu_{0}M\Delta t}.\\  
\end{split}
\end{equation}
Finally, the mass variance $\sigma_\delta$ during the matter-dominated era can be calculated as follows \cite{green2004NewCalculation}
\begin{equation}\label{sigmadelta}
    \sigma_\delta^2 =\frac{1}{2}\int^\infty_0 W^2(\Tilde{k}_\chi,\mathcal{R})\left( \frac{\Tilde{k}_\chi}{{k}_\chi}\right)^4 \mathcal{P}_{\zeta_\text{n-ad}}\frac{d\Tilde{k}_\chi}{\Tilde{k}_\chi},
\end{equation}
where we introduced the window function $W( \Tilde{k}_\chi,\mathcal{R})=\exp \left(  -\Tilde{k}_\chi^2 \mathcal{R}^2/2 \right)$, with $\mathcal{R}=1/k_\psi$.
Finally,  we can use  $\sigma_\delta$ calculated from Eqs.\eqref{sigmadelta}, \eqref{beta} and \eqref{sigmadelta008} to obtain the constraint on PBH production to be
\begin{equation}\label{sigmafinal}
       0.0755^2\gtrsim \frac{0.08318  \cdot e^{0.6992 \cdot M\Delta t}}{(M\Delta t)^{3/2}}\cdot   \frac{m^4_\chi k_\chi^3}{\sqrt{\xi}M^5 M_p^2}
\end{equation}

\begin{table}[t]
    \centering
    \begin{tabular}{|c|c|c|c|}
    \hline
       $m_\chi$    &  $k_\chi$  & $\Delta t  $ ($ M_p^{-1}$)\\
         \hline
         
        $10^{-12}M_p$   & $10^{-12}M_p$ &  $2.24 \times 10^7$  \\
        \hline
        $10^{-12}M_p$    & $10^{-10}M_p$ &  $2.02 \times 10^7$  \\
        \hline
        $10^{-12}M_p$     & $10^{-8}M_p$ &  $1.80 \times 10^7$  \\
        \hline
        $10^{-8}M_p$     & $10^{-12}M_p$ &  $1.65 \times 10^7$   \\
        \hline
        $10^{-8}M_p$     & $10^{-10}M_p$ &  $1.43 \times 10^7$   \\
        \hline
        $10^{-8}M_p$     & $10^{-8}M_p$ &  $1.21 \times 10^7$  \\
        \hline
        $10^{-6}M_p$     & $10^{-12}M_p$ &  $1.36 \times 10^7$   \\
        \hline
        $10^{-6}M_p$     & $10^{-10}M_p$ &  $1.136 \times 10^7$   \\
        \hline
        $10^{-6}M_p$     & $10^{-8}M_p$ &  $9.12 \times 10^6$  \\
        \hline
    \end{tabular}
    \caption{The duration of the required PBH abundance. In this table, we used $\xi=1$. Based on the table, $\Delta t$ is insensitive to the chosen parameters.}
    \label{tablepbh}
\end{table}

Eq. \eqref{sigmafinal} shows the dependence on $k_\chi$, $\xi$, $m_\chi$, and $\Delta t$. In the following, we present the numerical results for Eq. \eqref{sigmafinal} in Table \ref{tablepbh}. From our numerical analysis, by varying $k_\chi$ and $m_\chi$ with fixed $\xi$, it appears that $\Delta t$ is insensitive to the chosen parameters. The duration required to obtain a sufficient amount of PBHs is found to be $\sim 10^{7}M_p^{-1}$. This value is consistent with Fig. \ref{kksmall}, which shows that the effective growth of $\chi$ production occurs after $\Delta t > 10^{6}M_p^{-1}$. Note that in our numerical calculation, we keep $m_\chi, k_\chi \ll \xi M^2$. In this case, the PBH constraint may not jeopardize our model. Moreover, even if $\chi$ grows larger, at some point it cannot collapse into PBHs due to the rapidly expanding universe. Thus, PBHs are suppressed beyond a certain stage \cite{green2001primordial}. 

Lastly, in calculating the PBH abundance in this section, we employed the semi-analytical calculation developed in refs. \cite{green2001primordial,liddle2000super}, which was found to differ quantitatively from the numerical results (see Fig. \ref{analytic-numeric}). However, as discussed above, we have verified that this discrepancy affects the preheating duration, $\Delta t$, only logarithmically, as shown in eq. \ref{sigmafinal}. In addition, the calculation of refs. \cite{green2001primordial,liddle2000super} obtains the minimum duration of PBH production by setting $\mu_p$ to its maximum value. This choice yields the highest possible PBH production during preheating. As we show later in subsection \ref{reheatbypreheat}, the PBH duration is constrained by the reheating temperature; thus, a longer duration is strongly disfavored.

\section{Reheating scenario}\label{reheatingscenario}
In this section, we will discuss the possible reheating scenario in this model. It is noted that the reheating stage is the condition where the cosmological model can be closely linked to SM particle physics. However, it turns out that calculating the reheating temperature in \ref{secreheating} is not straightforward, but using the e-fold connection disregards the actual mechanism. We will see in the following subsection that several mechanisms can appear, and we will discuss which mechanism is the most favored.

\subsection{The perturbative decay of the scalaron to scalar \texorpdfstring{$\chi$}{chi}}
The most traditional depiction of the reheating temperature is based on the simple perturbative decay of the inflaton into light and relativistic particles. It is assumed that the energy of these light particles is converted into radiation. If we consider the scalaron decaying to $\chi$ via the Lagrangian
\begin{equation}
 \frac{\phi}{\sqrt{6}M_p} g^{\mu\nu}\partial_\mu \chi \partial_\nu \chi,
\end{equation}
the decay rate, within approximation $m_{\chi}\ll M$, can be written as
\begin{equation}
\Gamma(\phi \rightarrow \chi\chi) \simeq \frac{1}{384\pi} \frac{M^3}{M_p^2},
\end{equation}
and the reheating temperature can be easily evaluated as
\begin{equation}
T_\text{reh} = \left( \frac{90}{g^* \pi^2} \right)^{1/4} \sqrt{M_p \Gamma},
\end{equation}
where $g^* \sim 100$ is the SM degrees of freedom. As $M$ is well-constrained by ACT, it is clear that the reheating temperature can be obtained as
\begin{equation}
T_\text{reh} \approx 3 \times 10^9  \text{GeV}.
\end{equation}
However, this value is 5 orders of magnitude smaller than our prediction after applying the ACT constraint. In this case, the reheating temperature from this perturbative mode is strongly discouraged.

\subsection{The perturbative decay of the scalaron to fermion \texorpdfstring{$\psi$}{psi}}

As previously remarked, the perturbative decay of the scalaron to $\chi$ is heavily suppressed by ACT results. In that case, the interaction between the scalaron and fermions is described by the Dirac Lagrangian in the Einstein frame, which reads
\begin{equation}\label{diracL}
\mathcal{L}_\text{Dirac} = e^{-2\sqrt{\frac{2}{3}}\frac{\phi}{M_p}}\Bar{\psi} (i\gamma^\mu \partial_\mu - m_\psi)\psi,
\end{equation}
where $\Bar{\psi}$ and $\psi$ are antifermion–fermion pairs, $\gamma^\mu$ is the Dirac gamma matrix, and $m_\psi$ is the mass of the fermions. Due to the kinetic term of the Dirac Lagrangian being on-shell, the interaction term of the scalaron with $\Bar{\psi}\psi$ is controlled by
\begin{equation}
2\sqrt{\frac{2}{3}} \frac{m_\psi}{M_p} \phi \Bar{\psi}\psi.
\end{equation}
Straightforwardly, we obtain the decay rate as
\begin{equation}
\Gamma(\phi \rightarrow \Bar{\psi}\psi) \simeq \frac{1}{3\pi} \frac{m_\psi^2 M}{M_p^2},
\end{equation}
and the reheating temperature as high as
\begin{equation}
T_\text{reh} \simeq 5.3 \times 10^{-4}m_\psi \sqrt{1-\frac{4m_\psi^2}{M^2}}.
\end{equation}
By using the ACT reheating temperature constraint of $\sim 10^{14}$ GeV, $m_\psi$ should be of the order of the Planck mass $M_p$, which is kinematically forbidden ($m_\psi\gg M$). Thus, this reheating mechanism is also disfavored.

\subsection{Remarks on the perturbative mode}\label{remarks}
 In our discussion of perturbative reheating via the scalar field $\chi$ and fermion $\psi$, we show that the perturbative decay of both scalar and fermion modes is disfavored by the ACT results, as it yields a low reheating temperature.

In ref. \cite{allahverdi2010reheating}, it is shown that both reheating mechanisms are rather oversimplified, as they neglect the violation of the fluctuation--dissipation theorem (see Refs. \cite{hu1995fluctuation,berera2005absence,gleiser1994microphysical}). Additionally, these mechanisms overlook the effects of preheating and can only be significant if preheating is inefficient. 
In short, the realistic mechanism considered in our paper can be summarized as follows: At the first stage of preheating, the inflaton experiences its first zero crossing and produces massive non-thermal $\chi$ particles. In our model, the production of fermions $\psi$ is suppressed by Pauli blocking; hence, immediately after inflation, the Universe is filled predominantly with massive $\chi$ particles. These massive $\chi$ particles do not decay into daughter fields, as they redshift immediately.
After several oscillations, when the Universe is close to a radiation-dominated regime, the inflaton oscillation still occurs as a narrow resonance. The last burst of particle production due to the inflaton zero crossing then determines the reheating temperature. During this stage, the $\chi$ particles decay immediately into relativistic fermions $\psi$, which subsequently reheat the Universe. This assumption has been used extensively in Refs. \cite{risdianto2025second,risdianto2025inflation}, and it will be discussed in more detail in Sec. \ref{reheatbypreheat}.

\subsection{Reheating temperature by preheating}\label{reheatbypreheat}
As mentioned in the previous subsection, the reheating temperature should be evaluated by the last burst of the preheating stage before the radiation-dominated regime, and calculate the reheating mechanism based on the assumption made in \ref{remarks}.

To calculate the reheating temperature. Firstly, we need to write the relation between the energy density $\delta \rho$ produced by the single crossing during the 'last' preheating and the reheating temperature $T_\text{reh}$ via 
\begin{equation}
    T_\text{reh}=\left(\frac{30}{g^*\pi^2} \delta \rho\right)^{1/4}.
\end{equation}
Explicitly,  $\delta \rho$ can be evaluated as \cite{risdianto2025second,hashimoto2021inflation}

\begin{equation}\label{deltarhofinal}
    \delta \rho=\int^{\pi M^{-1}}_0 dt  \hspace{1mm}\Gamma_{\chi\rightarrow \Bar{\psi}\psi} n_\chi\Bar{m}_\chi  e^{-\int^t_0\Gamma_{\chi\rightarrow \Bar{\psi}\psi}dt'},
\end{equation}
where\footnote{Assuming the Floquet exponent varies slowly over the dominant instability band, we approximate
$\mu_k \simeq \mu_0$  for modes  $k \lesssim k_*$.
The occupation number therefore grows as $n_k \sim e^{2\mu_0 Mt}$. }
\begin{equation}\label{nchi}
    n_\chi=\int \frac{d^3k_\chi}{(2\pi)^3}e^{2\mu_p M \Delta t} \simeq \frac{k_\chi^3}{6\pi^2}e^{2\mu_0 M\Delta t}
\end{equation}
denotes the number density of $\chi$, and we use 
\begin{displaymath}
    e^{2\mu_0 M\Delta t}=e^{2\left(\frac{\ln 3}{2\pi}\right)M\Delta t}=3^{M\Delta t/\pi}.
\end{displaymath}
From the PBH analysis in section \ref{pbh}, we found that the duration of preheating required to produce a sufficient abundance of PBHs is of the order $\Delta t\sim10^7M_p^{-1}$. Since both PBH production and reheating originate from the same preheating process, it is natural to assume that the duration relevant for reheating is at least of the same order of magnitude. As will be shown below, the duration of preheating plays a crucial role in determining the resulting reheating temperature through the Floquet enhancement factor. Consequently, significantly larger values of $\Delta t$ would quantitatively modify the resulting reheating temperature.
In that case, we may obtain $3^{M\Delta t/\pi} \approx 10^{48}$. 

In addition, the mass $\Bar{m}_\chi$ is estimated from the Eq. \eqref{chioscillation} as 
\begin{equation}
    |\Bar{m}_\chi|=\sqrt{\sqrt{6} \frac{\xi M^2\Tilde{\phi}}{M_p}\sin(Mt)}.
\end{equation}
To complete Eq. \eqref{deltarhofinal}, we should consider that there is an interaction between $\chi$ and $\psi$ via 
\begin{equation}
    -y\chi \Bar{\psi}\psi,
\end{equation}
where $y$ is the Yukawa coupling between $\chi$ and $\Bar{\psi}\psi$.
The decay rate for this channel is approximately given by,
\begin{equation}
    \Gamma_{\chi\rightarrow\Bar{\psi}\psi}\simeq \frac{y^2m_\chi}{8\pi}.
\end{equation}
With these, we can solve the Eq. \eqref{deltarhofinal} as
\begin{equation}
    \delta \rho \simeq 3.71 \times 10^{42}\, y^2\, k_\chi^3\, m_\chi\, \sqrt{\xi}  
\end{equation}
Straightforwardly, the reheating temperature can be estimated as
\begin{equation}
    T_\text{reh}\simeq 5.8 \times 10^{10}\left(y^2 k_\chi^3   m_\chi\sqrt{\xi}   \right)^{1/4}.
\end{equation}
The reheating temperature is now shown to be dependent on several parameters. In short, one can see the Table. \ref{table2}. It is shown the required parameters $y$, $k_\chi$, and $m_\chi$ to achieve the favored reheating temperature of $\sim 10^{14}$ GeV.

\begin{table}[t]
    \centering
    \begin{tabular}{|c|c|c|}
    \hline
        $y$  &  $m_\chi$   & $T_\text{reh}$\\
    \hline
    \hline
        $10^{-12}$  &   $10^{3} $ GeV&  $1.1 \times 10^{14}$ GeV  \\
    \hline
        $10^{-13}$  &   $10^{4} $ GeV&  $6.36 \times 10^{13}$ GeV  \\
    \hline
        $10^{-14}$  &   $10^{6} $ GeV &  $6.36 \times 10^{13}$ GeV  \\
    \hline
        $10^{-15}$  &  $10^{8} $ GeV &  $6.36 \times 10^{13} $ GeV  \\ 
    \hline
        $10^{-16}$  &  $10^{10} $ GeV &  $6.36 \times 10^{13}$ GeV  \\
    \hline
         $10^{-17}$  &  $10^{12}$ GeV &  $6.36 \times 10^{13}$ GeV  \\ 
    \hline
    \end{tabular}
    \caption{The reheating temperature with varying $y$ and $m_\chi$. We set $k_\chi=10^{-7}M_p$ and  $\xi=1$ and adjust the parameters to be in the range of $T_\text{reh}=10^{13} \sim 10^{14}$ GeV.}
    \label{table2}
\end{table}

Considering the chosen parameters, we need to clarify some favored values. For $m_\chi$, it should be higher than the electroweak (EW) scale, as $\chi$ and fermion $\psi$ have bare masses, and their masses do not come from the symmetry breaking, they should not belong to the SM particles\footnote{As it is noted that the SM particles are massless at higher than the EW scale} and both should lie above the EW scale. On the other hand, bounded from above, $m_\chi\lesssim M/2$, to ensure the perturbativity of the inflaton decay. $\xi$ is constrained by unitarity, so it is safe to set $\xi=1$. Lastly, during the narrow resonance, $k_\chi$ should be higher than $\sqrt{M^3/M_p}$ (see Eq. \eqref{chioscillation}) to ensure the transition from matter to radiation-dominated, which favors relativistic particle production. We assumed that during the early stage of preheating, the particle spectrum is dominated by infrared (IR) modes generated through resonant amplification. As the system evolves, nonlinear rescattering and backreaction broaden the spectrum and progressively populate ultraviolet (UV) modes, thereby transferring energy to higher momenta $k_\chi$. Thus, the momentum distribution evolves throughout preheating, with infrared $k_\chi$ modes dominating at early times and higher-momentum modes becoming increasingly populated at later stages.

Lastly, Table \ref{table2} presents the preferred parameter values that accommodate the constrained reheating temperature. From the table, it can be seen that a reheating temperature of approximately $10^{14}$ GeV can be achieved. However, the assumption of preheating duration can be vital in this calculation. This issue may be discussed in future works \cite{Risdianto2026PreheatingStarobinsky}. We emphasize that the parameter choices adopted in the present work should be regarded as phenomenologically motivated benchmark values within the effective field theory of our model. In particular, the small Yukawa coupling $y\lesssim10^{-12}$, the initial spectator field value $\chi(0)\sim10^{-3}M_p$, and the masses $(m_\chi,m_\psi)$ are chosen to simultaneously achieve successful reheating, PBH production, and consistency with the observational constraints considered in this work. The question of whether these parameter values can be naturally realized within a UV-complete theory lies beyond the scope of the present work.

\section{Conclusion}\label{conclusion}
In this paper, we revisit Starobinsky inflation, with particular focus on preheating and reheating mechanisms in light of recent ACT results, especially the ACT-LB data. These new observations place tighter constraints on the allowed parameter space of the model through the relation between the number of e-folds and the reheating temperature. In particular, achieving a sufficiently high reheating temperature favors an e-fold number in the range  $N \simeq 55\text{--}56 $ with $T_\text{reh} \sim 10^{14}$ GeV, allowing Starobinsky inflation to remain consistent with the updated observational constraints. Consequently, the reheating mechanisms must be modified to align with these new results. Such a high reheating temperature affects the entire reheating mechanism related to the preheating stage. In addition, the direct perturbative decay after inflation from the scalaron to scalar $\chi$ is strongly disfavored if the preheating mechanism is dominant and the higher reheating temperature is expected. On the other hand, reheating via the decay of the scalaron to fermions $\psi$ is more severely disfavored. The higher reheating temperature obtained in this work may have important implications for the subsequent thermal history of the Universe, including thermal particle production and other cosmological observables. A detailed investigation of these effects requires the detailed particle physics model and solving the corresponding Boltzmann equations, which is beyond the scope of the present work.

For the efficient preheating to occur, a certain initial condition of $\chi$ with $\chi(0)\gtrsim 10^{-3}M_P$ needs to exist at the start of preheating. This initial condition is believed to have existed because $\chi$ is a spectator field during inflation. The fluctuation of this spectator's field during inflation is beyond the scope of this paper.  

Ultimately, the reheating mechanism is realized through a specific preheating process. During resonance, the oscillating scalaron produces the scalar $\chi$. While fermion production is also possible, it is heavily suppressed due to the Pauli exclusion principle. During the preheating stage, perturbative decay of the scalaron and $\chi$ occurs but proceeds very slowly and does not contribute adequately to the reheating temperature. At the end of the preheating stage, the final burst of particle production from the scalaron to $\chi$, followed by the immediate decay into fermions $\psi$, becomes the sole contributor to the reheating temperature. The reheating mechanism via this mode is found to be consistent with our chosen parameters.

\begin{table}[t]
    \centering
    \begin{tabular}{|c|c|c|}
     \hline
     Parameter & value & Note\\
    \hline
      $y$   &  $ \lesssim 10^{-12}$ &  reheating temperature constraint \\
    \hline
      $\chi(0)$ & $\sim 10^{-3}M_p$ & successful preheating \\
    \hline
        $\xi$ & $ \lesssim 10$ & unitarity issue\\
    \hline
       $m_\chi$ & $10^{-6}\sim 10^{-14}M_p$ &  \shortstack{ $m_\chi< \frac{1}{2} M$ (no inflaton's remnant condition), \\
       $m_\chi\gg m_\psi$ (successful reheating scenario)\\and
       $m_\chi\gg 10^2$ GeV  (EW scale)}\\
    \hline
      $m_\psi$ & $10^{-7}\sim 10^{-15}M_p$ & \shortstack{$m_\psi\ll m_\chi$ (successful reheating scenario)\\ and $m_\psi\gg 10^2$ GeV (EW scale)}\\
    \hline
    \shortstack{$k_\chi$ mode\\ (early preheating)} & $ \ll 10^{-11}M_p^2$&\shortstack{ $\ll  \xi M^2$ \\ (efficient preheating at the end of inflation)}\\
    \hline
    \shortstack{$k_\chi$ mode \\ (end of preheating)} & $ \gg 10^{-7}M_p^2$ & \shortstack{$\gg \sqrt{M^3/M_p}$ \\ (relativistic particle production)}\\
    \hline
    \end{tabular}
    \caption{The viable values of the parameters in our model.}
    \label{table3}
\end{table}

In the following, we consider the favored parameters of our model based on the discussion in the previous sections. We investigate whether these parameters are compatible with PBH production during preheating. In our model, $y$ is constrained by the reheating temperature to be $\lesssim 10^{-12}$. The non-minimal coupling $\xi$ should naturally not exceed $10$ to avoid unitarity issues. Furthermore, $m_\chi$ should be lower than $\frac{1}{2}M$ to ensure the complete decay of the scalaron, and larger than $m_\psi$ and the electroweak (EW) scale ($\gtrsim 10^3$~GeV $\approx 10^{-14}M_p$), as $\chi$ must differ from SM particles to preserve the viability of our model\footnote{If $m_\chi$ were at the EW scale and $\chi$ were not the Higgs field, it might already have been observed. Moreover, its mass does not originate from symmetry breaking, which further indicates that it should not be an SM particle.}. The mass $m_\psi$ should be much lower than $m_\chi$ to ensure a successful reheating scenario\footnote{Since $\psi$ is expected to reheat the Universe, it should be relativistic. Thus, $m_\psi\ll m_\chi$ is preferred.}. During the early stage of preheating, the particle spectrum is dominated by low-momentum modes. As preheating progresses, backreaction and rescattering effects gradually populate higher-momentum states. Although the comoving Fourier label $k_\chi$ is fixed for each mode, the occupation-number distribution evolves dynamically over time. Finally, Table~\ref{table3} shows the preferred parameters used in this paper.

Finally, we emphasize that the parameter choices adopted in this paper should be regarded as phenomenologically motivated benchmark values within the effective field theory of our model. In particular, the small Yukawa coupling, the initial spectator field value, and the masses $(m_\chi,m_\psi)$ are chosen to simultaneously achieve successful reheating, PBH production, and consistency with the observational constraints considered in this work. The origin of these parameter values from a UV-complete theory lies beyond the scope of the present work.

\section{Acknowledgment}
M.A.S. would like to thank the Indonesia Endowment Fund for Education (LPDP) Scholarship Program, Ministry of Finance, Republic of Indonesia.

%\appendix

%\begin{thebibliography}{99}
\bibliographystyle{JHEP}
\bibliography{name}

%\end{thebibliography}
\end{document}